\documentclass{jetpl}

\usepackage[english]{babel}
\usepackage{graphics}
\twocolumn

\newcommand\PW{82mm}

\title{Evolution of One-Dimensional Wind-Driven Sea Spectra}

\rtitle{Evolution of One-dimensional}

\sodtitle{Evolution of One-Dimensional}

\author{A.\,I.\,Dyachenko$^{1,2,}$\/\thanks{alexd@itp.ac.ru},
        D.\,I.\,Kachulin$^{1}$, V.\,E.\,Zakharov$^{1,2,3,4}$}

\rauthor{A.\,I.\,Dyachenko$^{1,2}$, D.\,I.\,Kachulin$^{1}$, V.\,E.\,Zakharov$^{1,2,3,4}$}

\sodauthor{Dyachenko, Kachulin, Zakharov}

\address{$^1$Novosibirsk State University, 630090, Novosibirsk-90, Russia\\~\\
$^2$Landau Institute for Theoretical Physics, 142432, Chernogolovka, Russia\\~\\
$^3$Department of Mathematics, University of Arizona, Tucson, AZ, 857201, USA\\~\\
$^4$Physical Institute of RAS, Leninskiy prospekt, 53, Moscow, 119991, Russia}

\dates{ \ }{*}

\abstract{
We analyze modern operational models of wind wave prediction on the subject for compliance dissipation. Our numerical simulations from the "first principle" demonstrate that heuristic formulas for damping rate of free wind sea due 
to "white capping" (or wave breaking) dramatically exaggerates the role of this effect in these models.
}

\begin{document}

\maketitle


\section{Introduction}

We perform numerical simulation of evolution of surface waves spectra that has been excited by wind. 
One of the motivation of writing this article is purely practical. Measure of nonlinearity of wave at the surface of deep water is their average steepness $\mu = <\nabla\eta^2>$, where $\eta(\vec r,t)$ is the shape of the surface. Characteristic
value of $\mu$ in real sea is moderate, $\mu \simeq 0.06-0.07$. However, even at small steepness "white capping" (or 
wave breaking) happens occasionally, due to what waves loose energy. This phenomenon is not studied yet, either 
experimentally nor theoretically.
Nevertheless in the operational models of wind waves prediction heuristics formulas  for rate of wave decay (due to this 
phenomenon) are widely used. They were introduced about thirty years before \cite{Komen84,WAMDI88} and little has changed 
since then. In our opinion they have no serious justification. The goal of this article is to check these heuristic formulas 
by numerical experiments not assuming statistical description.

To study "white capping" model with one horizontal dimension is enough. If steepness is moderate, $\mu \leq 0.07$, one can 
use dynamical "Zakharov equation" \cite{Z68}, which is greatly simplified in 1-D case. It reduces to the simple Hamiltonian 
system which is very convenient for numerical simulation \cite{DZ2011-1,DZ2012-2}. Canonical transformation resulting to 
this model is described in detail in \cite{DKZ2016}. In the framework of this model we perform numerical simulations
for very long time (hundreds of thousands of characteristic wave period) and make sure that heuristic formulas 
\cite{Komen84,WAMDI88} give to large rates of energy decay. It makes to treat used below wave prediction operational models
highly critical.

Another motivation for this work is the desire to describe (possible more in detail) phenomenon of "white capping" for the 
waves with so moderate steepness. This work is not finished yet, but we established a most important fact - wave breaking 
is preceded by "freak wave" which actually breaks. "Freak waves" appear naturally as a result of modulational instability
\cite{Z68}, but even stable spectra of moderate amplitude are able to generate them. Although freak waves are now rare events 
separated by time interval it tens of thousands of wave periods.

Finally, in this article we come back to the old question about integrability of the free surface hydrodynamics of the deep 
water. Hypothesis of integrability was formulated in the paper \cite{DZ94} the result which was a key when deriving compact 
equation \cite{DZ2011-1,DZ2012-2}. Since then it was argued both against integrability \cite{DLZ95,DKZ2013} and in favor
integrability \cite{ZD00,DZ08}. In our experiments we observed behavior which is typical for integrable systems. Dynamics of the 
wave field was quasi-periodical. Spectra averaged over great time ((of the order of hundred thousands wave periods) have 
changed little, loosing 15\% of their energy due to arising rare "freak waves".

As an initial condition we used experimental and often cited in the oceanographic literature JONSWAP spectrum \cite{Hass73} 
with the wind speed 12m/sec.

\section{JONSWAP Spectrum}

\begin{figure}
\includegraphics[angle=-00,width=\PW]{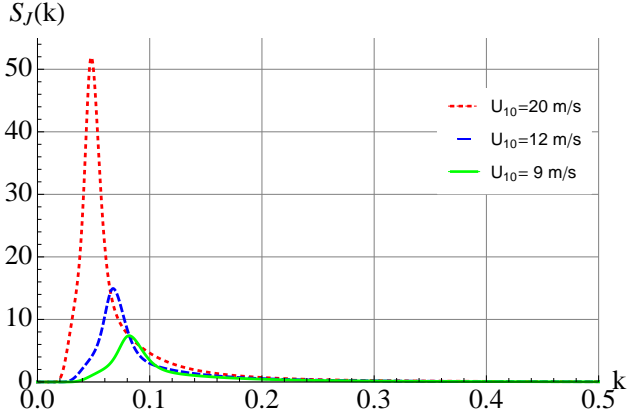}%
\caption{{\bf Fig.1} Energy density for JONSWAP spectrum for different wind speeds. }
 \label{FIG_JONSWAP}
\end{figure}

Hasselmann et al, in \cite{Hass73}, have analyzed data collected during the Joint North Sea Wave Observation Project - JONSWAP, and found that the wave spectrum is never fully developed. It continues to develop through non-linear, wave-wave interactions even for very long times and distances. They therefore proposed a spectrum in the form:
\begin{eqnarray}\label{JS}
S_J(\omega)d\omega &=& \alpha\frac{g^2}{\omega^5}\mbox{exp}^{\left [ -\frac{5}{4}(\frac{\omega_0}{\omega})^4\right ]} \gamma^{r}d\omega,\cr
r &=& \mbox{exp}^{\left [ -\frac{(\omega - \omega_p)^2}{2\sigma^2\omega_p^2})\right ]}
\end{eqnarray}
\begin{eqnarray}\nonumber
\alpha &=& 0.076(\frac{U_{10}^2}{F g})^{0.22},\cr
\omega_p &=& 22\left ( \frac{g^2}{U_{10} F}\right )^{\frac{1}{3}}, \cr
\gamma &=& 3.3, \cr
\sigma &=& \begin{cases} 
0.07  \mbox{ if } \omega \leq  \omega_p
\cr 0.09   \mbox{ if } \omega \leq  \omega_p
\end{cases}
\end{eqnarray}
Here $U_{10}$ - wind speed at the altitude 10m, $F$ - is fetch, e.i. distance from the shore.
Spectra with different wind seeps are shown in Figure \ref{FIG_JONSWAP}.

In the present article we study the relaxation of this developed sea state.

\section{Compact equation for water waves}

We start with well known Hamiltonian for water waves
\begin{multline} \label{Ham4}
H = \frac{1}{2}\int {g\eta^2 + \psi \hat k\psi} dx 
    -\frac{1}{2}\int \{(\hat k\psi)^2 -(\psi_x)^2\}\eta dx +\\
+\frac{1}{2}\int \{\psi_{xx} \eta^2 \hat k\psi + \psi \hat k(\eta \hat k(\eta \hat k\psi))\} dx + \dots
\end{multline}
which is expanded up to the fourth order as a function of Hamiltonian variables $\eta$ and $\psi$ (see \cite{Z68}):
After introducing complex canonical variables $a_k$
$$
\eta_k\! =\! \sqrt{\frac{\omega_k}{2g}}(a_k+a^*_{-k}) \hspace{.2cm}
\psi_k \!=\! -i\sqrt{\frac{g}{2\omega_k}}(a_k-a^*_{-k}) \hspace{1em}
\omega_k = \sqrt{gk}
$$
in the articles \cite{DZ2011-1,DZ2012-2} we applied canonical transformation to the Hamiltonian variable $a_k$ to introduce normal canonical variable $b(x,t)$:
\begin{enumerate}
\item $\eta_k, \psi_k \Rightarrow \mbox{normal canonical variable } a_k$
\item $a_k \Rightarrow b_k$
\end{enumerate}
This transformation explicitly uses vanishing of four-wave interaction and possibility to consider surface waves moving in the 
same direction, see \cite{DZ2011-1,DZ2012-2}. For this variable $b(x,t)$ Hamiltonian (\ref{Ham4}) acquires nice and 
elegant form:
\begin{multline}\label{SPACE_NICE}
{\cal H} =  \int\!b^*\hat\omega_k bdx + \\
+ \frac{1}{2}\int\!\left |\frac{\partial b}{\partial x}\right |^2
\left [\frac{i}{2}\left ( b \frac{\partial b^*}{\partial x} - b^*\frac{\partial b}{\partial x}\right ) -\hat K|b|^2 \right ] dx.
\end{multline}

Corresponding equation of motion is the following:
\begin{multline}\label{MotionSPACE}
i\frac{\partial b}{\partial t} = \hat\omega_k b
+\frac{i}{4}\hat P^+\left [ b^* \frac{\partial}{\partial x} ({b'}^2) -
\frac{\partial}{\partial x}( {b^*}' \frac{\partial}{\partial x}b^2) \right ]- \\
-\frac{1}{2} \hat P^+ \left [ b \cdot \hat k(|b'|^2) - \frac{\partial}{\partial x}(b'\hat k (|b|^2))\right ],
\end{multline}
Eigenvalue of the projection operator $\hat P^+$  in the Fourier-space is  step-function:
\begin{equation}\nonumber
P^+_k=\theta(k) =\left\{%
\begin{array}{ll}
    1, & k>0; \\
    0, & k\leq 0. \\
\end{array}%
\right.
\end{equation}
Transformation from $b(x,t)$ to physical variables $\eta(x,t)$ and $\psi(x,t)$ can be recovered from canonical transformation. It has been derived in \cite{DKZ2016}. Here we write this transformation up to the second order:
\begin{multline}\label{EPX_up_2_2order}
\eta(x) = \frac{1}{\sqrt{2}g^{\frac{1}{4}}}(\hat k^{\frac{1}{4}}b(x)+\hat k^{\frac{1}{4}}b(x)^*)+\frac{\hat k}{4\sqrt{g}}[\hat k^{\frac{1}{4}}b(x) - \hat k^{\frac{1}{4}}b^*(x)]^2,\\
\psi(x) = -i\frac{g^{\frac{1}{4}}}{\sqrt{2}}(\hat k^{-\frac{1}{4}}b(x)-\hat k^{-\frac{1}{4}}b(x)^*)+ \\
+\frac{i}{2}[\hat k^{\frac{1}{4}}b^*(x)\hat k^{\frac{3}{4}}b^*(x) - 
\hat k^{\frac{1}{4}}b(x)\hat k^{\frac{3}{4}}b(x)] + \cr
+\frac{1}{2}\hat H[\hat k^{\frac{1}{4}}b(x)\hat k^{\frac{3}{4}}b^*(x)+
\hat k^{\frac{1}{4}}b^*(x)\hat k^{\frac{3}{4}}b(x)].
\end{multline}
Here $\hat H$ - is Hilbert transformation with eigenvalue $i\textbf{sign}(k)$.

\section{Kinetic equation}

Along with simulation in the framework of equation (\ref{MotionSPACE}) we have been solving the same initial problem with simple quasi-linear model
\begin{equation}\label{FP}
\frac{\partial |b_k|^2}{\partial t} = -\gamma_{diss}|b_k|^2 .
\end{equation}
performing averaging by time and wave numbers, so that 
\begin{equation}\nonumber
|b_k|^2 \rightarrow n_k = <|b_k|^2>,
\end{equation}
{\it here expressions for $\gamma_{diss}$ are taken from the article \cite{WAMDI88} (these are formulas 2.10 and 2.16), namely}:
\begin{eqnarray}\label{W34}
\gamma_1^{WAM3} &=& 3.33\cdot 10^{-5}\bar\omega(\frac{\omega}{\bar\omega})^2(\frac{\bar\alpha}{\bar\alpha_{PM}})^2,\cr
\gamma_2^{WAM3} &=& 2.33\cdot 10^{-5}\hat\omega(\frac{\omega}{\hat\omega})^2(\frac{\hat\alpha}{\hat\alpha_{PM}})^2.
\end{eqnarray}
Here $\bar\omega$ and $\hat\omega$  mean averaging over spectrum, $\alpha$ is an integral wave steepness, and 
$$
\bar\alpha_{PM} = E\bar\omega^4 g^{-2},
$$
$$
\bar\alpha_{PM} = 4.57\cdot 10^{-3}
$$
is the theoretical value of $\bar\alpha$ for a Pierson-Moskowitz spectrum \cite{PM64}, and
$$
\hat\alpha_{PM} = 0.66\cdot\bar\alpha_{PM}.
$$
$E$ - is the total energy (surface elevation variance). {\it Again, all these definitions are taken from operational 
models from \cite{Komen84,WAMDI88}}. More recent models have just slight corrections to them.
In this case model (\ref{FP}) is equivalent to the well-known Hasselmann kinetic equation \cite{HASS1962} (see also 
\cite{DLZ95}) because wind pumping is absent and collision term $S_{nl}$ (due to the result of \cite{DZ94}) identically 
equal to zero. In both models we add artificial damping 
\begin{equation}\label{GD}
\Gamma_d(k) = \begin{cases} 
\alpha k^4 & \mbox{ if highest harmonics of } b_k \mbox{ are $10^4$ times} 
\\  & \mbox{ greater then roundoff errors} 
\\ 0 &  \mbox{ in the other case}
\end{cases}
\end{equation}
with $\alpha = 0.9 / \tau k_{max}^4$. It provides dissipation of extreme waves due to wave breaking.
We calculated effective damping due to wave breaking, $<\gamma_{diss}>$, plugging results of calculations in the framework of (\ref{MotionSPACE}) into the equation (\ref{FP}). Another words we define  $<\gamma_{diss}>$ as following:
\begin{equation}\label{FP1}
\gamma_{diss} = -\frac{1}{|b_k|^2}\frac{\partial |b_k|^2}{\partial t}.
\end{equation}

\section{Evolution of the JONSWAP spectrum}

We study relaxation of developed sea with different wind speeds - $U_{10} = 9m/sec$, $U_{10} = 12m/sec$ and
$U_{10} = 20m/sec$. However in this article we show results of simulation for  $U_{10} = 12m/sec$ only. 
The others are very similar.
Periodic domain of the length $L=10000$ meters was used for numerical simulations.
Initial conditions for $b_k$ where chosen according to JONSWAP spectrum:
\begin{equation}\label{bk0}
|b_k|^2=\sqrt{\frac{2g}{\omega_k}} |\eta_k|^2 = S_J(k)\frac{2\pi}{L}\frac{g}{\omega_k}.
\end{equation}
Phases of $b_k$ were chosen randomly in the interval $[0:2\pi]$.
Fetch $F$ was equal to 157000 meters.

\begin{figure}
\includegraphics[angle=-00,width=\PW]{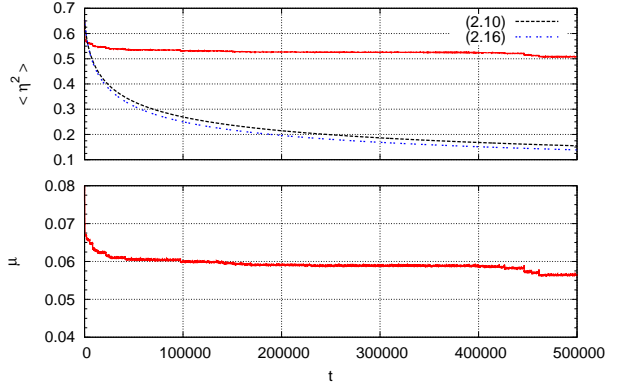}%
\caption{{\bf Fig.2} Energy density and steepness for the wind 12 m/sec. Time is measured in seconds.}
\label{FIG_02}
\end{figure}

We observed much smaller dissipation than predicts WAM3 model. For the wind velocity $U_{10}=12 m/sec$ energy density both in our numeric experiment and calculated according to \cite{Komen84,WAMDI88} are shown in Figures \ref{FIG_02}.
Energy density is measured in oceanographic units 
$$
\frac{\mbox{energy density}}{g}=\mbox{meter}^2.
$$
The average steepness $\mu$ is calculated as following:
$$
\mu = \sqrt{\int_{-\infty}^\infty k^2 |\eta_k|^2 dk}.
$$
In the picture one can see initial fast relaxation of energy in numerical experiment. It is due to dissipation of long tail 
$\simeq \omega^{-5}$ of JONSWAP spectrum in k-space (see (\ref{JS})). After initial relaxation there are rare events of energy dissipation in our experiment. Average steepness is also shown in the Figure \ref{FIG_02}.

\begin{figure}
\includegraphics[angle=-00,width=\PW]{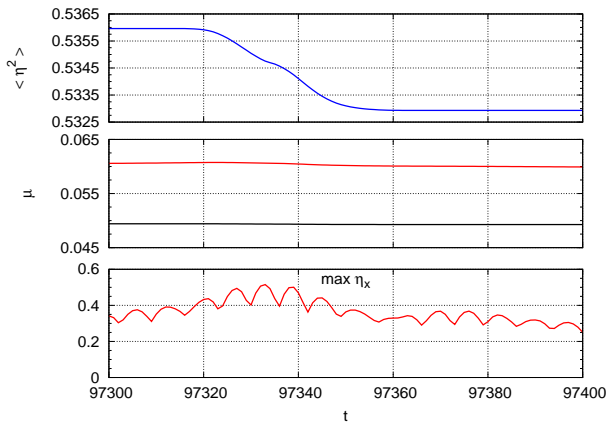}%
\caption{{\bf Fig.3} Drop of energy due to extreme wave appearing (wave breaking). Last picture shows maximal steepness of the extreme wave.}
 \label{FIG_03}
\end{figure}

One of this rare events, wave breaking, taking place at $\mbox{time}\simeq 93340$, is shown in detail in the Figure \ref{FIG_03}. One can see oscillation of the amplitude of the extreme wave.    

\begin{figure}
\includegraphics[angle=-00,width=\PW]{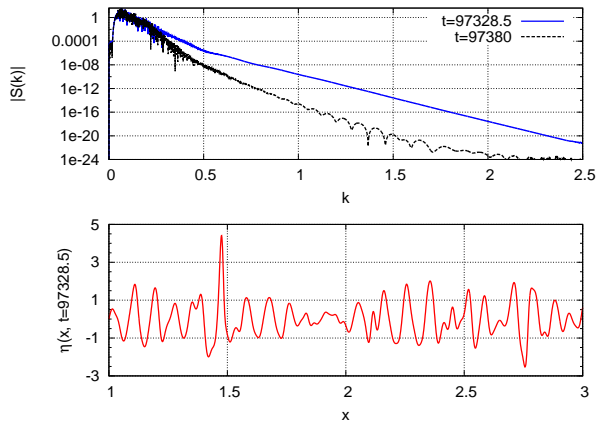}%
\caption{{\bf Fig.4} Spectral density $S(k)$ at the moment of freak wave appearing and freak wave almost 5 meters height.}
 \label{FIG_04}
\end{figure}

Spectrum $S(k)$ along with zoomed profile of the surface at $\mbox{time}\simeq 93340$ is shown in the Figure \ref{FIG_04}.
One can see the amplitude of the extreme (freak) wave more that 3 times large then for nearby waves. 
Also in the Figure \ref{FIG_04} energy spectrum $S(k)$ after wave breaking is shown. It does not have tail in large wavenumbers.

\begin{figure}
\includegraphics[angle=-00,width=\PW]{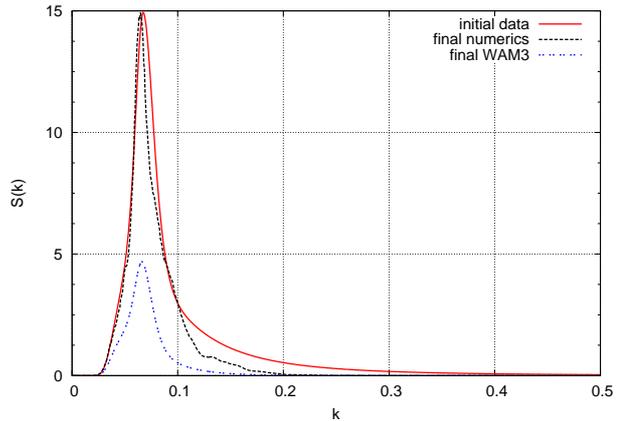}%
\caption{{\bf Fig.5} Spectral density $S(k)$ at initial moment (solid line), final numerical spectrum (dashed line) and final WAM3 spectrum (double dotted line).}
 \label{FIG_05}
\end{figure}

Great difference between numerical results and prediction of WAM3 model is seen in Figure \ref{FIG_05}.
Both of them had the same initial condition. However at the final time the spectra are very different. WAM3 predicts much more energy dissipation. It is also seen in Figure \ref{FIG_02}.

\begin{figure}
\includegraphics[angle=-00,width=\PW]{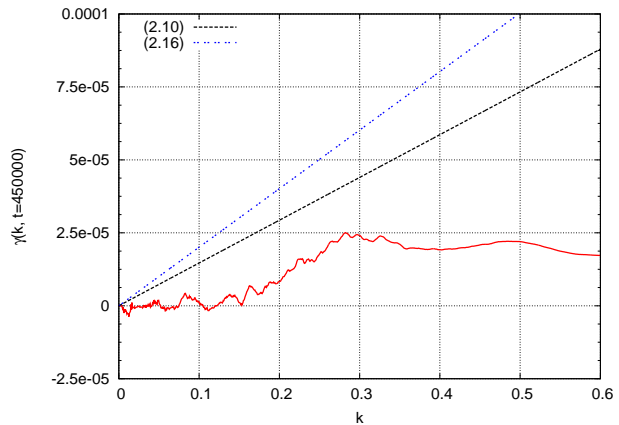}%
\caption{{\bf Fig.6} Compare $\gamma_1^{WAM3}$, $\gamma_2^{WAM3}$ and  $<\gamma_{diss}>$ }
 \label{FIG_06}
\end{figure}

One can see that relaxation of energy is sufficiently long process. During hundreds of thousands seconds it 
decreases by $\simeq 20\%$. During this time we calculated average $<\gamma_{diss}>$ according to (\ref{FP1}).
To make it smooth enough time of averaging was few hours (10000 sec). In the Figure \ref{FIG_06} there are dissipations
according to (\ref{W34}) {\it (or equations (2.10) and (2.16) in \cite{WAMDI88})}  plotted by dotted and double dotted lines. $<\gamma_{diss}>$ 
calculated with the use of dynamical equation with (\ref{GD}) shown solid line.
One can see that numerical experiment gives much less value of dissipation. Moreover, dissipation is absent in the core
of spectral density where $k_0\simeq 0.06-0.07 m^{-1}$.

\section{Conclusion}

The main result of our work is the fact that heuristic formulas for damping rate of free wind sea (\ref{W34}) due 
to "white capping" dramatically exaggerates the role of this effect. Especially convincing is Figure \ref{FIG_06} 
showing that in the region of spectral maximum dissipation of energy is practically absent. Increase of $<\gamma_{diss}>$ 
with increasing of wave number indicates that damping is concentrated in the region of large wave numbers. it means that 
"white capping" leads primarily to vanishing of the spectra "tails" and smoothing of the wave field. We stress that our
simulations describe sea evolution during few days after "switch off" wind. During this time sea lost no more then 20\%
of the energy. Similar picture of slow energy dissipation was observed in \cite{ZKP09}. Because "dissipation function" 
$\gamma_{diss}$ plays a key role in the massively used operational models the inevitable conclusion is that these models need 
to be fundamentally reviewed.

Our simulations are another argument in favor of integrability of deep water hydrodynamics. Others arguments in this favor 
are given in \cite{ZD00} and are very serious. There it is shown that exact system of Euler equations describing potential 
flow of deep water with a free surface can have any number of commuted integrals of motion. Weak point of this argument is the question about completeness of tis system of integrals. In the article \cite{DKZ2013} it is shown that model 
(\ref{MotionSPACE}) is not integrable. But nonintegrability arises in the fifth order of the perturbation theory where 
equation (\ref{MotionSPACE}) strictly speaking is not applicable. The most serious arguments contained in \cite{DLZ95}, 
where indicated the non-existence of higher integrals.

\section{Acknowledgments}

This work was supported by the Grant "Wave turbulence: theory, numerical simulation, experiment" \#14-22-00174 of Russian Science Foundation.

Numerical simulation was performed on the Informational Computational Center of the Novosibirsk State University.

\end{document}